\documentclass[conference]{IEEEtran}
\usepackage{graphicx,cite,calc,epstopdf}
\usepackage{tikz}
\usetikzlibrary{shapes,arrows}
\usepackage{rotating}
\usepackage{amsmath}
\begin{document}
%\addtolength{\voffset}{0.6cm}
%
%\title{Data Efficient Spatial Monitoring of Field Signals Using UAV: A Stochastic Gradient Approach}
\title{E-CONDOR: Efficient Contour-Based Detection Of Random Spatial Signals From UAV Observations Using Dual Stochastic Gradient}
%\title{---------------------------------------------------------------------------------------------------------------------------------------}

\author{\IEEEauthorblockN{Maryam Zahra, Homa Tajiani and Hadi Alasti}
\IEEEauthorblockA{Purdue University, Fort Wayne (PFW)\\
       %\IEEEauthorblockA{School of Polytechnic\\Purdue University, Fort Wayne (PFW)\\
Fort Wayne, IN, 46805 USA\\
Email: mzahra@pfw.edu, htajiani@purdue.edu, halasti@purdue.edu}
          %Email: ----------------------------------------------------------}
}

\maketitle

\begin{abstract}
This paper presents a novel efficient method for spatial monitoring of the distribution of correlated field signals, such as temperature, humidity, etc. using unmanned aerial vehicles (UAVs). The spatial signal is compressed to its iso-contour lines at a number of known levels that are introduced by data fusion center (DFC). The UAV traces a contour line of the field signal at a time, and reports the coordinates of its own traces to the DFC for spatial modeling. The DFC iteratively improves the spatial model of the field signal and assigns a new contour level to each UAV to trace and report its coordinates for spatial model improvement. The selected batch of levels and the start point of the search are introduced by the DFC.
In order to reduce the required data for spatial modeling, and accordingly improve the algorithm’s data efficiency, dual stochastic gradient routines are used at the DFC to find a next proper number of contour levels in the batch, and to eliminate the redundant contour levels, in each iteration. The performance evaluation of the proposed algorithm based on computer simulations demonstrates significantly faster convergence, better signal estimation, and a higher data efficiency against when the stochastic gradient is not used.
\end{abstract}

% For peer review papers, you can put extra information on the cover
\IEEEpeerreviewmaketitle

\section{Introduction}
\begin{figure}[t!]
	\centering
		\includegraphics[width=3.4in,angle=0]{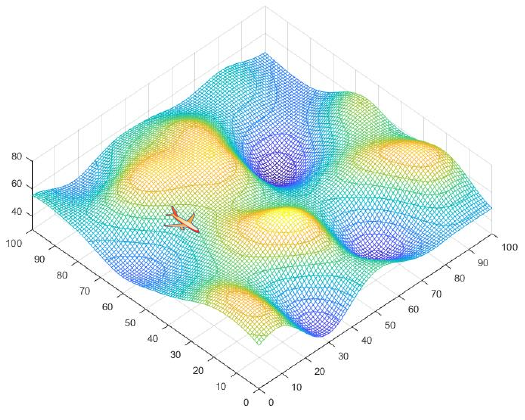}
		\caption{The spatial distribution is compressed into its contour lines, and a UAV reports its coordinate when tracing the contour lines at specific levels.}
		\label{fig: 01}
\end{figure}

This paper presents a data efficient algorithm for spatial monitoring of correlated field signals such as temperature, humidity, gas density, radiation intensity, etc. using unmanned aerial vehicle (UAV), similar to the illustrated spatial distribution in Fig.~\ref{fig: 01}. For high data efficiency, the spatial signal is compressed to its contour lines at a number of known levels; and stochastic gradient descent is used for selection of proper batches of contour levels. As a machine learning algorithm, stochastic gradient (SG) was used for spatial monitoring using wireless sensor observations~\cite{Hadi-SG-2019, Hadi-WSS-2020, Hadi-ITNAC-2021, Hadi-UEMCON-2021,Hadi-mdpi-2021}. 
The advantage of the proposed dual SG are saving the algorithm's cost, improving its general convergence, and data efficiency, as results of these two improvements.
Machine learning algorithms were used for spatial object recognition~\cite{Seo,Gao,Tang,Zhang}, signal characterization~\cite{Hussein,Lv,Chaki,Alzahed,Wickramasinghe,Hamilton}, image processing~\cite{Asokan,Silva,Adel,Sun,Sivanathan}, etc. The proposed data efficient approach in this paper is based on the accelerated learning~\cite{Hadi-UEMCON-2021, Hadi-mdpi-2021}, where it reduces the volume of the required data by dropping the number of iterations.
Similar to level crossing sampling for one-dimensional signals~\cite{Hadi-IET-WSS,Hadi-IGI}, the unknown spatial signal distribution is compressed into a batch of its $M$ contour lines at levels $\{\ell_k\}_{k=1}^M$, as previously addressed in~\cite{Hadi-ITNAC-2021, Hadi-UEMCON-2021,Hadi-mdpi-2021,Hadi-SG-2019,Hadi-WSS-2020,Hadi-MISS-2017,Alasti_FOWANC}. For spatial monitoring based on contours lines, it is required to detect or trace the contour lines. The detection of contours and edge of a threshold using static wireless sensors were reported in several researches, among them in ~\cite{Hadi-GreenCom,Armstrong,Gandhi,Suri,Sarkar,Lian}. Unlike them, in this paper, we investigate a dual SG algorithm as a machine learning approach to efficiently model a distributed spatial signal. The number of contour levels iteratively increases, until tangible convergence of the algorithm. The number of required contour levels is estimated using one SG process. Next, for higher data efficiency, a second SG algorithm drops a portion of redundant contour levels. Most of the computation is performed at DFC, where the UAV(s) only trace the ordered contour lines of the spatial signal, and report the paced trace after compression to the DFC. 

The DFC uses mean-absolute-error (MAE) as a measure for performance evaluation of spatial modeling. The total paced length of the contour lines is used as the cost of the algorithm, that also represents the volume of the collected data from the field. In this research, we assume that the effect of noise in UAV observations is negligible. Also, we ignored the flying distance between two distinct contour pieces.

In the upcoming section of this paper, the problem and the solution statements are given. Next, the background of the problem is reviewed. Later, the proposed dual SG algorithm is detailed. Finally, the performance of the proposed algorithm is evaluated and discussed. 

\section{The Problem and The Solution Statement}
An unknown correlated spatial signal $g(x,y)$, similar to Fig.~\ref{fig: 01} is distributed over a given known area $A$. The strength of the signal is measured closely using one or a few UAVs. The spatial signal is modeled using its $M$ iso-contour lines at levels $\{\ell_k\}_{k=1}^M$. It is assumed that during the multi-contour line observation, the spatial signal remains unchanged. However, it may change gradually during a temporal phase after the spatial observation is done. The objective is to \textit{efficiently monitor the spatial signal} based on contour line observations. Here, the UAV(s) do not have any initial information about the number of contour lines $M$, and their levels $\{\ell_k\}_{k=1}^M$.

To solve the problem, each UAV communicates with a computationally powerful data fusion center (DFC). The DFC estimates the spatial signal based on the reported coordinates of the signal strengths. The DFC computes a batch of contour levels and sequentially broadcasts the levels to the UAV(s) in the signal field. To efficiently monitor the spatial signal, the DFC uses stochastic gradient algorithms that are among the known machine learning algorithms.

\section{The Related Background}
In this section and before detailing the dual SG algorithm, here we review the technical background of the machine learning algorithm for spatiotemporal monitoring of correlated signals using static wireless sensors. 

Efficient spatiotemporal monitoring of a signal distribution over a wide area using the observations of a large number of wireless sensors ($N_0$) was discussed in~\cite{Hadi-SG-2019, Hadi-WSS-2020, Hadi-ITNAC-2021, Hadi-UEMCON-2021,Hadi-mdpi-2021}. The spatial signal is compressed into its $M$ contour lines at known levels $\{\ell_k\}_{k=1}^M$. Only $N_r$ wireless sensors whose observations fall within the $\Delta$ margin of each of the introduced $M$ contour levels report to the DFC. $N_r$ is estimated in (\ref{equ: 01}), where $f_g(s)$ is the probability density function (PDF) of the signal strength.
\begin{equation}
N_r = N_0 \sum_{k=1}^{M}\int_{\ell_k-\Delta}^{\ell_k+\Delta}f_g(\gamma)d\gamma \cong 2 N_0 \Delta \sum_{k=1}^{M} f_g(\ell_k)
\label{equ: 01}
\end{equation}
The DFC iteratively improves the monitoring performance by increasing the number of contour levels $M$. To drop the number of reporting sensors, the $\Delta$ margin shrinks based on SG algorithm, according to (\ref{equ: 02}), as discussed in~\cite{Hadi-SG-2019}. The introduced algorithm uses the apriori contour-based model to update the spatial model, over time.
\begin{equation}
\Delta_n = \Delta_{n-1}(1 +\frac{\nabla Error_{n-1}}{2\overline{Error}_{n-1}})
\label{equ: 02}
\end{equation}
Here, $\nabla{Error_{n-1}}=(Error_{n-1} - Error_{n-2})$ and $\overline{Error}_{n-1} = 1/2(Error_{n-1} + Error_{n-2})$. To make the gradient $\nabla{Error_{n-1}}$ independent from the instantaneous error, it is normalized to the instantaneous learning mean-error $\overline{Error}_{n-1}$ is ~\cite{Bershad}, where the learning error of the $n^{th}$ iteration is computed according to (\ref{equ: 03}), and $\tilde{g}_n(x_i,~y_j)$ is the spatial signal estimation in the $n^{th}$ iteration step, at $P \times Q$ grid points $(x_i,~y_j)$, based on the collected contour line details.
\begin{equation}
Error_n = \sum_{i=1}^P\sum_{j=1}^Q\frac{\left|\tilde{g}_n(x_i,~y_j)-\tilde{g}_{n-1}(x_i,~y_j)\right|}{P\times Q}
\label{equ: 03}
\end{equation}
To accelerate the speed of convergence, an additional SG algorithm was introduced in~\cite{Hadi-UEMCON-2021} to offer a proper $M_n \leftarrow M_{n-1} + \kappa_n$ for the number of contour levels in the next iteration. The level-increment factor $\kappa_n$, rises gradually as the number of iterations $n$ increases. 
The performance evaluation results proved that the introduced algorithm is faster, meaning that it converges in less iterations, and has higher data efficiency. 

Unfortunately, spatial monitoring using a large number of static wireless sensors is not realistic and feasible. Accordingly, spatial monitoring using UAV observations is explored based on the developed research contributions in static wireless sensor fields.
\section{The Proposed Dual Stochastic Gradient Learning Algorithm} 
In this section the dual SG algorithm is introduced, based on the reviewed background. The algorithm consists of two parallel SG processes. The first process computes the number of level-increments were it results in finding the new batch of contour levels. The second process, however, drops a selected redundant number of contour levels from the most recent batch that are too close to at least one of the previously paced contour levels.
\subsection{Initiation Phase}
\begin{figure}[b]
  \centering
	\includegraphics[width=3.5in,angle=0]{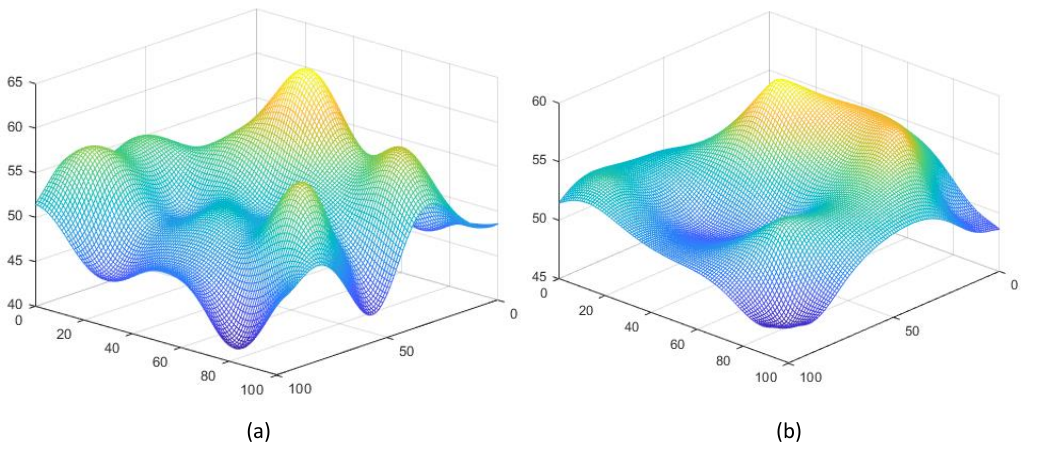}
  \caption{The original spatial signal: (a) is estimated based on a first rough spatial estimation: (b), after the initiation phase}
	\label{fig: 02}	
\end{figure}

To initiate the spatial monitoring algorithm, the UAV(s) fly over very limited, arbitrary routes and report its coordinates and the strength of the spatial signal to the DFC. The DFC estimates the spatial signal based on the received information. In this research, we assume that the UAV(s) fly over the two main diagonal routes, and the DFC provides a first estimation of the spatial signal from the UAV reports.
Fig.~\ref{fig: 02} illustrates a random correlated spatial signal distribution over an area $A$. In initiation phase, a rough estimation of the spatial signal is made based on limited and arbitrary paced routes by the UAV(s).

\subsection{Iterative Phases of the Algorithm}
Having received the initiation phase data of the spatial signal, the DFC reconstructs the spatial signal using bipolar spline interpolator that is detailed in~\cite{Sandwell}. Then, the DFC computes the signal strength range, and subsequently a batch of contour levels $\{\ell_k\}_{k=1}^M$, for the given $M$. The smallest $M$ in the first round of iteration and after \textit{initiation phase} is $3$. In this research we assume optimally spaced levels based on Lloyd-Max algorithm~\cite{Sayood}. The optimally spaced levels are defined in (\ref{equ: 04}) and (\ref{equ: 05}). The Lloyd-Max levels are used for optimal quantization of a signal at $M$ levels, in order to minimize the modeling error. It is worthy to mention that the $M$ levels are defined based on the spatial signal strength's range, which is computed after the most recent spatial signal estimation. 
\begin{equation}
\ell_i = \frac{\int_{y_i}^{y_{i+1}}xf_g(x)dx}{\int_{y_i}^{y_{i+1}}f_g(x)dx}, ~~~~i=1,2,\cdots, M
\label{equ: 04}
\end{equation}
and $y_i$ in according to (\ref{equ: 05}).
\begin{equation}
y_i = \frac{\ell_i + \ell_{i-1}}{2}, ~~~~ i = 1, 2, \cdots, M-1
\label{equ: 05}
\end{equation}

After finding the $M$ levels, the DFC notifies the UAV(s) of their assigned level and an initiation coordinate $(x,y)$ to start pacing the contour lines of the distributed spatial signal in the field. Because the assigned coordinate is based on the most rough estimation of the recent spatial signal, instead of the actual spatial signal, the UAV may need to search for the actual initiation coordinate in the neighborhood of the assigned initiation coordinate $(x,y)$. 

Next, the UAV(s) pace the trace of each contour line and report the coordinates to the DFC for the next iteration of spatial modeling. Having assigned all the required number of contour levels to the UAV(s), and receiving the coordinates of the contour line traces, the DFC reconstructs the spatial signal and computes the learning error, as defined in~(\ref{equ: 03}). Then, the DFC modifies the level-increment $\kappa_n$ according to (\ref{equ: 06}); meaning that $M_n \leftarrow M_{n-1} + \kappa_n$. The learning error is used in the process update of each SG process according to (\ref{equ: 06}) and (\ref{equ: 07}). The $\lceil x \rceil$ operator in (\ref{equ: 06}) represents the smallest whole number larger than $x$ value.

\begin{equation}
\kappa_n = \kappa_{n-1} + \left\lceil1 + \frac{2|Error_{n-2} - Error_{n-1}|}{|Error_{n-2} + Error_{n-1}|}\right\rceil
\label{equ: 06}
\end{equation}
However, as $M$ increases, it is highly likely that a number of the computed contour levels for spatial signal observation in the next step be redundant, meaning that they are too close to the previously paced levels and there is no meaningfully new information in these redundant levels. Accordingly, we drop those redundant levels that are within $\delta_n$ neighborhood of any of the previously traced contour lines, and they are not assigned to any UAV to trace. $\delta_n$ is defined according to (\ref{equ: 07}). The recent contour level dropping policy results in reducing the algorithm's cost and increasing its data efficiency.
\begin{equation}
\delta_n = \delta_{n-1} |1 - \frac{2(Error_{n-2} - Error_{n-1})}{(Error_{n-2} + Error_{n-1})}|
\label{equ: 07}
\end{equation}
The iterative process of increasing the number of contour levels, tracing the contour lines by the UAV(s) and spatial signal estimation continues, until i) the learning error is tangibly dropped, ii) the spatial signal strength range converges. The summary of the proposed algorithm is tabulated in the following.
\begin{table}[h!]
\label{Tab: 01}
\noindent\textit{\textbf{The Summary of the proposed algorithm}}\\
\line(1,0){245}
\begin{enumerate}
	\item The DFC notifies the UAV(s) to pace arbitrary routes over the spatial signal field.
	\item The UAV(s) report the spatial signal strength at numerous coordinates to the DFC.
	\item The DFC makes a first estimation for the spatial signal strength: $\tilde{g}_0(x,y)$.
	\item The DFC computes the signal strength range from $\tilde{g}_0(x,y)$, and $M=3$ levels: $\{\ell_p\}_{p=1}^3$ (optimal levels are computed based on (\ref{equ: 04}) and (\ref{equ: 05})).
	\item The DFC estimates the contour lines of $\tilde{g}_0(x,y)$ for the computed levels, and finds an initiation coordinate $(x,y)$ for each contour piece.
	\item The DFC shares each level $\ell_p$ and an initiation coordinate $(x,y)$ with each UAV, in order.
	\item Each UAV finds the actual coordinate of the initiation coordinate $(x,y)$ of the contour line at level $\ell_p$.
	\item The UAV tracks its assigned contour line at level $\ell_p$ and reports its coordinates to the DFC for the next round of iteration: $n+1 \leftarrow n$.
	\item The DFC estimates the spatial signal distribution $\tilde{g}_{n+1}(x,y)$.
	\item The DFC computes the learning error $Error_{n+1}$ according to (\ref{equ: 03}).
	\item The DFC computes the $\delta_{n+1}$ and $\kappa_{n+1}$ according to (\ref{equ: 06}) and (\ref{equ: 07}).
	\item The DFC computes the new number of contour levels $M_{n+1} \leftarrow M_n + \kappa_{n+1}$ 
	\item The DFC computes the new batch of contour levels $\{\ell_p\}_{p=1}^{M_{n+1}}$ (optimally spaced based on (\ref{equ: 04}) and (\ref{equ: 05})).
	\item The DFC eliminates any of the redundant contour levels that are within the $\delta_{n+1}$ neighborhood of any of the past levels.
	\item The DFC finds the contour lines of the new spatial estimation $\tilde{g}_{n+1}(x,y)$ for the most recent batch of levels.
	\item Repeat the algorithm from the step (8), until convergence.
\end{enumerate}
\line(1,0){245}
\end{table}

\section{Performance Evaluation}
\subsection{Spatial signal model and assumption}
For performance evaluation of the introduced algorithm, first we review the spatial signal model. We use the diffusion model to model the spatial signal as previously discussed in ~\cite{Jindal_2}. The correlated spatial signal is randomly distributed over an area of \texttt{100 x 100}. For performance evaluation purpose using simulation, we use \texttt{MATLAB} and the related code is available online for verification purpose~\cite{UAV-1-codes-2024}. 
The synthetic correlated spatial signal $g(x,y)$ is generated using the analytical model according to~\ref{equ: 08}. According to this model, the spatial signal is generated by superposition of two groups of two-dimensional Gaussian distributions $G(mx, my, \sigma)$, according to (\ref{equ: 09}). Each group of Gaussian distribution has its own specific standard deviation of $\sigma = \sigma_1$ or $\sigma_2$. The center of the Gaussian distributions, $(mx_p, my_p)$ and $(\hat{mx}_p, \hat{my}_p)$ are random points inside the signal field. The coefficients of $a_p$ and $b_p$ in (\ref{equ: 08}) are positive random values, such that they limit the spatial signal strength's range within the range $(0,100)$. Fig.~\ref{fig: 02}-(a), represents the spatial signal, using (\ref{equ: 08}). For generation of this spatial signal, $\sigma_1$ and $\sigma_2$ are 10 and 15, respectively.
\begin{equation}
g(x,y) = \sum_{p=1}^{N_1} a_p G(mx_p, my_p, \sigma_1) + \sum_{p=1}^{N_2} b_p G(\hat{mx}_p, \hat{my}_p, \sigma_2)
\label{equ: 08}
\end{equation}
where the two dimensional Gaussian is according to (\ref{equ: 09}).
\begin{equation}
G(mx, my, \sigma) = exp(-\frac{(x-mx)^2+(y-my)^2}{2\sigma^2})
\label{equ: 09}
\end{equation}
\begin{figure}[b]
  \centering
	\includegraphics[width=3.2in,angle=0]{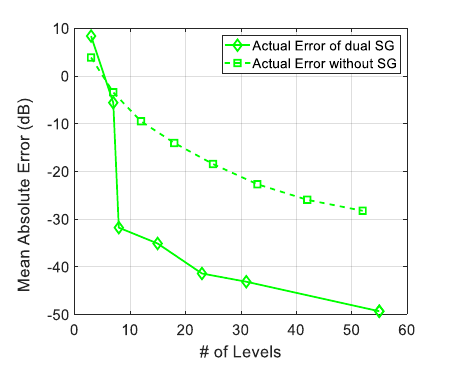}
  \caption{The spatial modeling error of Fig.~\ref{fig: 02} versus the number of levels. The figure compares the spatial modeling MAE in dB for: i) using the dual SG and ii) without dual SG.}
	\label{fig: 03}	
\end{figure}
In this performance evaluation we assumed UAV-based field observation. The mean absolute error (MAE) of the proposed algorithm is evaluated based on its cost, which is the length that the UAV(s) track. The MAE is defined according to (\ref{equ: 10}). In this performance evaluation we ignore the presence of observation noise.
\begin{equation}
E_n = \sum_{i=1}^P\sum_{j=1}^Q\frac{\left|g_n(x_i,~y_j)-\tilde{g}_n(x_i,~y_j)\right|}{P\times Q}
\label{equ: 10}
\end{equation}
\subsection{Performance evaluation results}
\subsubsection{Spatial modeling MAE}
The spatial modeling performance of the proposed algorithm (the dual SG) is compared with spatial monitoring without SG. The mean absolute error (MAE) as defined in (\ref{equ: 10}) is used for this purpose. Fig.~\ref{fig: 03}, represents the MAE of spatial modeling with and without dual SG, in dB, as defined in~(\ref{equ: 11}).  
\begin{equation}
\centering
	MAE_n~ (dB) = 20 Log_{10} (E_n)
\label{equ: 11}
\end{equation}
As this figure illustrates, spatial monitoring using the proposed dual SG algorithm converges faster and results in much lower modeling error than that of without dual SG.

\begin{figure}[t!]
  \centering
	\includegraphics[width=3.2in,angle=0]{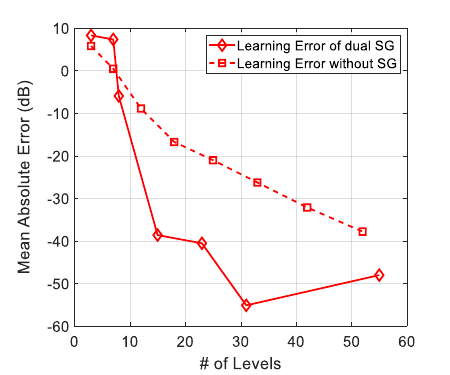}
  \caption{The spatial learning error of Fig.~\ref{fig: 02} versus the number of levels. The figure compares the spatial modeling MAE in dB for: i) using the dual SG and ii) without dual SG.}
	\label{fig: 04}	
\end{figure}

Fig.~\ref{fig: 04} represents the learning error of the dual SG algorithm, as defined in(\ref{equ: 03}). As this figure illustrates, the learning error of spatial modeling using the proposed dual SG algorithm converges faster and is much smaller than that of spatial modeling without dual SG.

\begin{figure}[b]
  \centering
	\includegraphics[width=3.1in,angle=0]{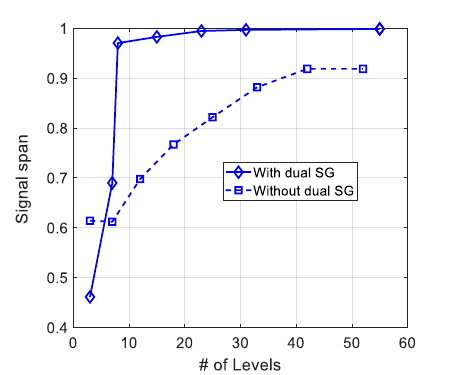}
  \caption{The figure compares the signal span ratio of the algorithm, with and without dual SG.}
	\label{fig: 05}	
\end{figure}
\subsubsection{Signal strength range span ratio}
During the spatial signal monitoring, the signal span range of the signal estimate converges to the signal span range of the actual signal. The signal span range is defined as: as: $spr_n = \frac{SP_n}{SP_{Actual}}$, where $SP_n$ is the signal strength range of the estimated spatial signal in the $n^{th}$ iteration; and $SP_{Actual}$ is the actual signal's span range, which is: the maximum signal strength - minimum signal strength. According to Fig.~\ref{fig: 05}, the dual SG algorithm converges faster and closer to 1, which is the ideal expected signal span after convergence. Also, the spatial modeling without dual SG converges slower, with meaningful bias.
\begin{figure}[h]
  \centering
	\includegraphics[width=3.1in,angle=0]{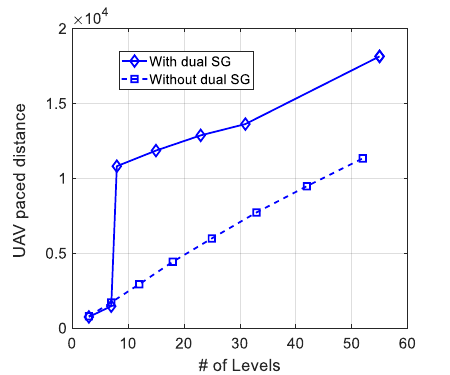}
  \caption{The figure compares the total flying distance of the UAV (the cost), with and without dual SG.}
	\label{fig: 06}	
\end{figure}
\begin{figure}[h!]
  \centering
	\includegraphics[width=3.1in,angle=0]{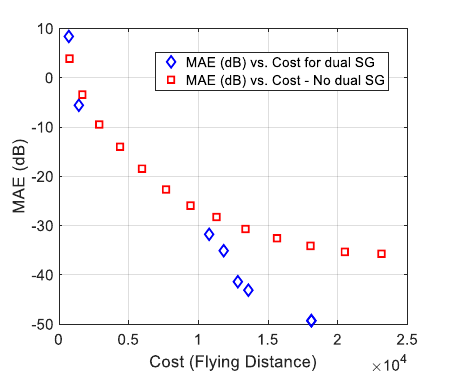}
  \caption{The spatial monitoring performance (MAE in dB), versus the total UAV flying distance (cost).}
	\label{fig: 07}	
\end{figure}
\subsubsection{The total flying distance (Cost)}
Fig.~\ref{fig: 06} represents the total flying distance of the spatial monitoring algorithms with and without dual SG. According to this figure, it looks like that for the same number of contour levels, the proposed dual SG algorithm is costlier than that of spatial monitoring without dual SG. However, by including the spatial monitoring performance (MAE) in this discussion we see that the proposed dual SG algorithm is less costly for a given target MAE. Fig.~\ref{fig: 07} illustrates the MAE in dB versus the total UAV flying distance. As this figure represents, the the proposed algorithm is tangibly less costly than that of spatial monitoring without dual SG, specifically for low amounts of MAE. This is one of the significance of the proposed algorithm.
\subsubsection{Variation of $\delta$ and the local spatial absolute error}
\begin{figure}[h!]
  \centering
	\includegraphics[width=3.1in,angle=0]{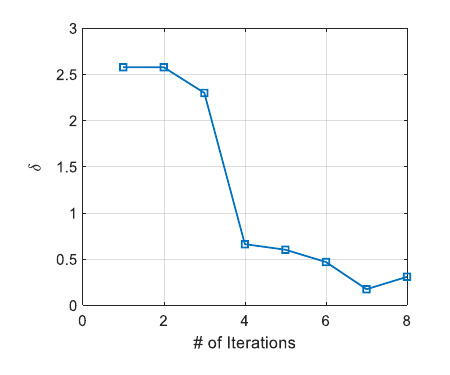}
  \caption{The figure represents the variation of $\delta$ against the iterations of the dual SG algorithm.}
	\label{fig: 08}	
\end{figure}

\begin{figure}[h!]
  \centering
	\includegraphics[width=3.3in,angle=0]{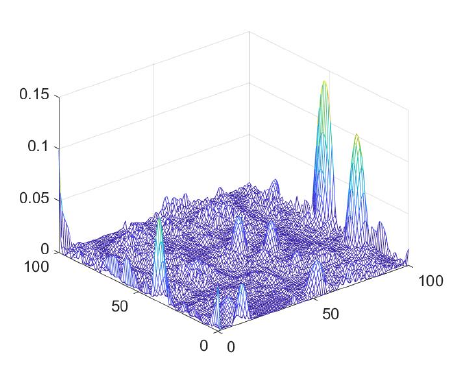}
  \caption{The figure represents the local spatial absolute error of monitoring Fig.~\ref{fig: 02}-(a) after 8 iterations, when dual SG is used.}
	\label{fig: 09}	
\end{figure}

The variation of the cost-reducer parameter $\delta$ also presents a dropping attribute is the sequence of the iterations of the algorithm, according to Fig.~\ref{fig: 08}. The noisy variations of this factor is related to the randomness in variation of $Error_n$, as defined in (\ref{equ: 03}). Fig.~\ref{fig: 09} represents the local spatial absolute error of monitoring Fig.~\ref{fig: 02}-(a) after 8 iterations, when dual SG is used. According to this figure, even the largest error magnitudes that appear around the borders of the area, are negligible.
\section{Conclusion}
Dual stochastic gradient (SG) is proposed for efficient monitoring of spatial signals using UAV-based observations. The spatial signal is modeled with a number of its iso-contour lines at known contour levels. Each UAV reports the coordinates of a given contour lines to data fusion center (DFC) for estimation of the spatial signal. The DFC improves the spatial monitoring performance by introducing some finer contour levels to the UAV(s) in the signal field. The dual SG is used to find the number of contour levels and to drop the redundant contour levels. The performance evaluation results shows that spatial monitoring using dual SG is significantly more efficient than that of without SG. In this study, the contour levels are chosen based on Lloyd-Max, which optimizes the spatial modeling. 
In the next step of this research, we will use the collected information from remote sensing sources to make it possible to track the temporal changes, efficiently.
%\section*{Acknowledgment}
%The author would like to appreciate the reviewers who improved this manuscript with their to-the-point comments.
%The authors would like to thank...

% that's all folks

\end{document}